\documentstyle{article}
\newtheorem{rem}{Remark}

\begin{document}
\begin{center}
\Large{\bf  TOWARDS  THE THEORY OF BENNEY EQUATIONS  \\[2mm]
}\vspace{4mm}\normalsize
\end{center}
 \begin{center}
{\bf Valery Dryuma}\vspace{2mm}\normalsize
\end{center}
\begin{center}
{\bf Institute of Mathematics and Informatics AS Moldova, Kishinev}\vspace{4mm}\normalsize
\end{center}
\begin{center}
{\bf  E-mail: valery@dryuma.com;\quad cainar@mail.md}\vspace{4mm}\normalsize
\end{center}
\begin{center}
{\bf  Abstract}\vspace{4mm}\normalsize
\end{center}

   The solutions of the Benney equations are
   constructed. Their properties are discussed.

\section{Introduction}

    A famous Benney equations describing a long waves on the surface of fluid has the form
\begin{equation}\label{dr1}
\frac{\partial f(x,v,t)}{\partial t}+v \frac{\partial f(x,v,t)
}{\partial x}-\frac{\partial A_0(x,t)}{\partial x}\frac{\partial
f(x,v,t)}{\partial v}=0,
\end{equation}
where $ A_0(x,t)=\int_{-\infty}^{+\infty}f(x,v,t)dv $

   It can be rewritten in the form of nonlinear system of p.d.e.
\begin{equation}\label{dr2}
\frac{\partial f(x,v,t)}{\partial t}+v \frac{\partial f(x,v,t)
}{\partial x}-\left(\frac{\partial g(x,v,t)}{\partial
x}+\frac{\partial h(x,v,t)}{\partial x}\right)\frac{\partial
f(x,v,t)}{\partial v}=0,$$$$ \frac{\partial g(x,v,t)}{\partial
v}=f(x,v,t),\quad\frac{\partial h(x,v,t)}{\partial v}=-f(x,v,t),
\end{equation}
where
\[
g(x,v,t)=\int_{-\infty}^v f(x,v,t) dv, \quad
h(x,v,t)=\int_{v}^{\infty} f(x,v,t) dv.
\]

   From the system (\ref{dr2}) we get the relations
\[
{\frac {\partial }{\partial x}}g(x,v,t)+{\frac {\partial
}{\partial x} }h(x,v,t)={\frac {{\frac {\partial }{\partial
t}}f(x,v,t)+v{\frac {
\partial }{\partial x}}f(x,v,t)}{{\frac {\partial }{\partial v}}f(x,v,
t)}}
\]
and
\[
\frac {\partial }{\partial v}g(x,v,t)+\frac {\partial }{\partial
v}h(x,v,t)=0,
\]
  which is equivalent  the nonlinear p.d.e.
\begin{equation} \label{dr3}
\left ({\frac {\partial }{\partial v}}f(x,v,t)\right ){\frac {
\partial ^{2}}{\partial t\partial v}}f(x,v,t)+\left ({\frac {\partial
}{\partial v}}f(x,v,t)\right ){\frac {\partial }{\partial
x}}f(x,v,t)+ \left ({\frac {\partial }{\partial v}}f(x,v,t)\right
)v{\frac {
\partial ^{2}}{\partial v\partial x}}f(x,v,t)-\]\[-\left ({\frac {\partial
^{2}}{\partial {v}^{2}}}f(x,v,t)\right ){\frac {\partial
}{\partial t} }f(x,v,t)-\left ({\frac {\partial ^{2}}{\partial
{v}^{2}}}f(x,v,t) \right )v{\frac {\partial }{\partial x}}f(x,v,t)
=0
 \end{equation}
 is followed.

    To construction of particular solutions of this equation  can be used the method of
    the (u,v)-transformation developed first by author .

\section{ The method of solution}

   To integrate the partial nonlinear differential equation
\begin{equation}\label{dryuma:eq11}
F(x,y,z,f_x,f_y,f_{xx},f_{xy},f_{yy},f_{xz},f_{yz},f_{zz})=0
\end{equation}
    can be applied a following method.

   We use the  change of the functions and  variables according
   to the rule
\begin{equation}\label{dryuma:eq12}
f(x,y,z)\rightarrow u(x,t,z),\quad y\rightarrow v(x,t,z),\quad
f_x\rightarrow u_x-\frac{v_x}{v_t}u_t,\quad f_z\rightarrow
u_z-\frac{v_z}{v_t}u_t,\quad f_y \rightarrow \frac{u_t}{v_t},...
\end{equation}

  In result instead of the equation (\ref{dryuma:eq11}) one get the
  relation between the new variables $u(x,t,z)$ and $v(x,t,z)$ and
  their partial derivatives
\begin{equation}\label{dryuma:eq13}
\Phi(u,v,u_x,u_t,u_z,v_x,v_t,v_z,...)=0.
  \end{equation}

    In some cases the integration of  the last equation is more simple problem than integration of  the equation
    (\ref{dryuma:eq11}).

    To illustrate this method let us consider some of examples.
\\[4mm]
    1.

    The equation
\begin{equation}\label{dryuma:eq14}
{\frac {\partial }{\partial x}}z(x,y)-\left ({\frac {\partial }{
\partial y}}z(x,y)\right )^{2}=0
\end{equation}
is transformed into the following form
\[
{\frac {\partial }{\partial x}}u(x,t)-{\frac {\left ({\frac
{\partial }{\partial t}}u(x,t)\right ){\frac {\partial }{\partial
x}}v(x,t)}{{ \frac {\partial }{\partial t}}v(x,t)}}-{\frac {\left
({\frac {
\partial }{\partial t}}u(x,t)\right )^{2}}{\left ({\frac {\partial }{
\partial t}}v(x,t)\right )^{2}}}=0.
\]

     Using the substitution
\[
u(x,t)=t{\frac {\partial }{\partial
t}}\omega(x,t)-\omega(x,t),\quad v(x,t)={\frac {\partial
}{\partial t}}\omega(x,t)
\]
we find the equation for $\omega(x,t)$
\[
{\frac {\partial }{\partial x}}\omega(x,t)+{t}^{2}=0.
\]

   Its integration lead to
   \[
   \omega(x,t)=-{t}^{2}x+{F_1}(t)
\]
where $F_1(t)$ is arbitrary function.

    Now with help of $\omega(x,t)$ we find the functions $u(x,t)$ and $v(x,t)$
\[
u(x,t)=-{t}^{2}x+t{\frac {d}{dt}}{F_1}(t)-{F_1}(t) ,\quad v(x,t)=
-2\,tx+{\frac {d}{dt}}{F_1}(t)
\]
or
\[
u(x,t)=ty+{t}^{2}x-{F_1}(t),\quad y= -2\,tx+{\frac
{d}{dt}}{F_1}(t).
\]

   After the choice of arbitrary function $F_1(t)$ and elimination of the parameter $t$ from these
   relations we get
   the function $z(x,y)$, satisfying the equation
   (\ref{dryuma:eq14}).

   2.

    The equation
\begin{equation}\label{dryuma:eq15}
{\frac {\partial }{\partial x}}z(x,y)+z(x,y)\left ({\frac
{\partial }{
\partial y}}z(x,y)\right )=0
\end{equation}
is transformed into the following form
\[{\frac {\partial }{\partial x}}u(x,t)-{\frac {\left ({\frac
{\partial }{\partial t}}u(x,t)\right ){\frac {\partial }{\partial
x}}v(x,t)}{{ \frac {\partial }{\partial t}}v(x,t)}}+u(x,t){\frac
{\left ({\frac {
\partial }{\partial t}}u(x,t)\right )}{\left ({\frac {\partial }{
\partial t}}v(x,t)\right )}}=0.
\]

     Using the substitution
\[
u(x,t)=t{\frac {\partial }{\partial
t}}\omega(x,t)-\omega(x,t),\quad v(x,t)={\frac {\partial
}{\partial t}}\omega(x,t)
\]
we find the equation for $\omega(x,t)$
\[
\left ({\frac {\partial }{\partial t}}\omega(x,t)\right ){t}^{2}-{
\frac {\partial }{\partial x}}\omega(x,t)-t\omega(x,t)=0.
\]

   Its integration give us the function
   \[
\omega(x,t)=t{F}\left({\frac {1-tx}{t}}\right)
\]
where $F()$ is arbitrary function.

    Now with help of the function $\omega(x,t)$ we can find the functions $u(x,t)$ and
    $v(x,t)$.  Then after the choice of arbitrary function $F()$ and elimination
of the parameter $t$ from the
   relations
\[
z=u(x,t),\quad y=v(x,t)
 \]
    we can get
   the function $z(x,y)$, satisfying the equation
   (\ref{dryuma:eq15}).

By analogy the substitution
\[
\left \{u(x,t)={\frac {\partial }{\partial t}}\omega(x,t),\quad
v(x,t)=t{ \frac {\partial }{\partial
t}}\omega(x,t)-\omega(x,t)\right \}
\]
into the equation for $\omega(x,t)$ lead to the equation
\[{\frac {\partial }{\partial x}}\phi(x,t)+{\frac {\partial }{\partial t
}}\phi(x,t)=0
\]
with general solution
\[
\phi(x,t)={F}(t-x)
\]
which also give us the solution of the equation
(\ref{dryuma:eq15}).

     3.

     The equation meeting in theory of the Benney equations
     \begin{equation}\label{dryuma:eq51}
     {\frac {\partial ^{2}}{\partial y\partial z}}f(x,y,z)+\left ({\frac {
\partial }{\partial y}}f(x,y,z)\right ){\frac {\partial ^{2}}{
\partial x\partial y}}f(x,y,z)-\left ({\frac {\partial }{\partial x}}f
(x,y,z)\right ){\frac {\partial ^{2}}{\partial {y}^{2}}}f(x,y,z)=0
\end{equation}
  after the (u,v)-transformation with the change $(y\rightarrow
t)$ takes the form
\[
\left ({\frac {\partial ^{2}}{\partial t\partial z}}u(x,t,z)\right
) \left ({\frac {\partial }{\partial t}}v(x,t,z)\right )^{2}-\left
({ \frac {\partial ^{2}}{\partial {t}^{2}}}u(x,t,z)\right )\left
({\frac {\partial }{\partial z}}v(x,t,z)\right ){\frac {\partial
}{\partial t} }v(x,t,z)+\]\[+\left ({\frac {\partial }{\partial
t}}u(x,t,z)\right )\left ({\frac {\partial }{\partial
z}}v(x,t,z)\right ){\frac {\partial ^{2}} {\partial
{t}^{2}}}v(x,t,z)-\left ({\frac {\partial }{\partial t}}u(x,
t,z)\right )\left ({\frac {\partial ^{2}}{\partial t\partial
z}}v(x,t, z)\right ){\frac {\partial }{\partial
t}}v(x,t,z)+\]\[+\left ({\frac {
\partial }{\partial t}}u(x,t,z)\right )\left ({\frac {\partial ^{2}}{
\partial t\partial x}}u(x,t,z)\right ){\frac {\partial }{\partial t}}v
(x,t,z)-\left ({\frac {\partial }{\partial t}}u(x,t,z)\right
)^{2}{ \frac {\partial ^{2}}{\partial t\partial
x}}v(x,t,z)-\]\[-\left ({\frac {
\partial }{\partial x}}u(x,t,z)\right )\left ({\frac {\partial }{
\partial t}}v(x,t,z)\right ){\frac {\partial ^{2}}{\partial {t}^{2}}}u
(x,t,z)+\left ({\frac {\partial }{\partial x}}u(x,t,z)\right
)\left ({ \frac {\partial }{\partial t}}u(x,t,z)\right ){\frac
{\partial ^{2}}{
\partial {t}^{2}}}v(x,t,z)=0.
\]

    In result of substitution of the form
\[
u(x,t,z)=t{\frac {\partial }{\partial
t}}\omega(x,t,z)-\omega(x,t,z),\quad  v(x,t,z)={\frac {\partial
}{\partial t}}\omega(x,t,z)
\]
we get from this relation the linear p.d.e. with the respect of
the function $\omega(x,t,z)$
\[
{\frac {\partial ^{2}}{\partial t\partial z}}\omega(x,t,z)+t{\frac
{
\partial ^{2}}{\partial t\partial x}}\omega(x,t,z)-{\frac {\partial }{
\partial x}}\omega(x,t,z)=0
\]

    Its solutions can be obtained by the Laplace method
    and after elimination of the parameter $t$ from the expressions for
    the function $(f(x,y,z)\rightarrow u(x,t,z))$
    and $(y\rightarrow v(x,t,z))$ the solutions of initial
    equation (\ref{dryuma:eq51}) can be constructed.

    By analogy this
method can be applied to obtaining particular solutions of the
equation (\ref{dr3}).

\section{The $(u,v)$-transformation of the equation (\ref{dr3})}

    For convenience we write the equation (\ref{dr3}) in the form
\begin{equation}\label{dr4}
\left ({\frac {\partial }{\partial y}}f(x,y,z)\right ){\frac {
\partial ^{2}}{\partial y\partial z}}f(x,y,z)+\left ({\frac {\partial
}{\partial y}}f(x,y,z)\right ){\frac {\partial }{\partial
x}}f(x,y,z)+ \left ({\frac {\partial }{\partial y}}f(x,y,z)\right
)y{\frac {
\partial ^{2}}{\partial x\partial y}}f(x,y,z)-$$$$-\left ({\frac {\partial
^{2}}{\partial {y}^{2}}}f(x,y,z)\right ){\frac {\partial
}{\partial z} }f(x,y,z)-\left ({\frac {\partial ^{2}}{\partial
{y}^{2}}}f(x,y,z) \right )y{\frac {\partial }{\partial
x}}f(x,y,z)=0,
\end{equation}
 where instead of the variables $v$ and
$t$ we used the variables $y$ and $z$.

   After the $(u,v)$-transformation with the change of the variable
   $y$ on parameter $t$ the equation (\ref{dr4}) takes the form of
   the relation between the functions $u(x,t,z)$ and $v(x,t,z)$
   and their derivatives
\[
\left ({\frac {\partial }{\partial t}}u(x,t,z)\right )\left
({\frac {
\partial ^{2}}{\partial t\partial z}}u(x,t,z)\right ){\frac {\partial
}{\partial t}}v(x,t,z)-\left ({\frac {\partial }{\partial
t}}u(x,t,z) \right )^{2}{\frac {\partial ^{2}}{\partial t\partial
z}}v(x,t,z)+ \]\[+\left ({\frac {\partial }{\partial
t}}u(x,t,z)\right )\left ({\frac {
\partial }{\partial t}}v(x,t,z)\right )^{2}{\frac {\partial }{
\partial x}}u(x,t,z)-\left ({\frac {\partial }{\partial t}}u(x,t,z)
\right )^{2}\left ({\frac {\partial }{\partial t}}v(x,t,z)\right
){ \frac {\partial }{\partial x}}v(x,t,z)+\]\[+\left ({\frac
{\partial }{
\partial t}}u(x,t,z)\right )v(x,t,z)\left ({\frac {\partial ^{2}}{
\partial t\partial x}}u(x,t,z)\right ){\frac {\partial }{\partial t}}v
(x,t,z)-\left ({\frac {\partial }{\partial t}}u(x,t,z)\right
)^{2}v(x, t,z){\frac {\partial ^{2}}{\partial t\partial
x}}v(x,t,z)-\]\[-\left ({ \frac {\partial ^{2}}{\partial
{t}^{2}}}u(x,t,z)\right )\left ({\frac {\partial }{\partial
t}}v(x,t,z)\right ){\frac {\partial }{\partial z} }u(x,t,z)+\left
({\frac {\partial }{\partial t}}u(x,t,z)\right )\left ({\frac
{\partial ^{2}}{\partial {t}^{2}}}v(x,t,z)\right ){\frac {
\partial }{\partial z}}u(x,t,z)-\]\[-v(x,t,z)\left ({\frac {\partial ^{2}}{
\partial {t}^{2}}}u(x,t,z)\right )\left ({\frac {\partial }{\partial t
}}v(x,t,z)\right ){\frac {\partial }{\partial x}}u(x,t,z)+v(x,t,z)
\left ({\frac {\partial }{\partial t}}u(x,t,z)\right )\left
({\frac {
\partial ^{2}}{\partial {t}^{2}}}v(x,t,z)\right ){\frac {\partial }{
\partial x}}u(x,t,z)
 =0.
\]

    There are a lot possibilities to bring this relation to one
    equation. Classification all types of reductions is open
    problem.

    We use a simplest type  of reductions.

    As example after the substitution
\begin{equation}\label{dr5}
u(x,t,z)=t{\frac {\partial }{\partial
t}}\omega(x,t,z)-\omega(x,t,z), \quad v(x,t,z)={\frac {\partial
}{\partial t}}\omega(x,t,z).
\end{equation}
this relation lead to the nonlinear partial differential equation
\begin{equation}\label{dryuma:eq16}
-t\left ({\frac {\partial }{\partial x}}\omega(x,t,z)\right
){\frac {
\partial ^{2}}{\partial {t}^{2}}}\omega(x,t,z)+\left ({\frac {
\partial }{\partial t}}\omega(x,t,z)\right ){\frac {\partial }{
\partial x}}\omega(x,t,z)-t{\frac {\partial ^{2}}{\partial t\partial z
}}\omega(x,t,z)+{\frac {\partial }{\partial
z}}\omega(x,t,z)-$$$$-\left ({ \frac {\partial }{\partial
t}}\omega(x,t,z)\right )t{\frac {\partial ^ {2}}{\partial
t\partial x}}\omega(x,t,z) =0.
\end{equation}

   Its particular solutions  can be used for
   construction of solutions of the equation (\ref{dr4}).

   Let as consider  some examples.

   Using  the substitution
\begin{equation}\label{dr6}
\omega(x,t,z)=A(t,z)+Bxt
\end{equation}
we get from the (\ref{dryuma:eq16}) the equation
\[
-{t}^{2}B{\frac {\partial ^{2}}{\partial {t}^{2}}}A(t,z)-t{\frac {
\partial ^{2}}{\partial t\partial z}}A(t,z)+{\frac {\partial }{
\partial z}}A(t,z)
=0
\]
with general solution
\begin{equation}\label{dryuma:eq17}
A(t,z)={\it \_F1}({\frac {zB-\ln (t)}{B}})+{\it \_F2}(z)t
\end{equation}
dependent from two arbitrary functions ${\it \_F2}(z)$ and  ${\it
\_F1}({\frac {zB-\ln (t)}{B}})$.

    In particular case
\[
{\it \_F1}(-{\frac {-zB+\ln (t)}{B}})={\frac {\left (-zB+\ln (t)
\right )^{2}}{{B}^{2}}}
\]
with  the help of the formulaes (\ref{dr5},~\ref{dr6}) we find the
relations
\[
f(x,y,z){B}^{2}+2\,zB-2\,\ln (t)+{z}^{2}{B}^{2}-2\,zB\ln (t)+\left
( \ln (t)\right )^{2} =0,
\]
\[
yt{B}^{2}+2\,zB-2\,\ln (t)-{\it \_F2}(z)t{B}^{2}-{B}^{3}xt=0.
\]

    From last equation we get the expression for parameter $t$
    \[
    t={e^{-{\it LambertW}(1/2\,{B}^{2}\left (-y+{\it \_F2}(z)+Bx\right ){e^{
zB}})+zB}}
\]
and after substitution its into the first one we find the function
$f(x,y,z)$
\[
f(x,y,z)=-{\frac {1+{\it LambertW}(1/2\,{B}^{2}\left (-y+{\it
\_F2}(z) +Bx\right ){e^{zB}})}{B}}
\]
which is solution of the equation (\ref{dr4}).

    Substitution
    \[
    v(x,t,z)=t{\frac {\partial }{\partial
    t}}\omega(x,t,z)-\omega(x,t,z),\quad
u(x,t,z)={\frac {\partial }{\partial t}}\omega(x,t,z)
\]
lead to the equation
\begin{equation}\label{dryuma:eq18}
t\left ({\frac {\partial }{\partial x}}\omega(x,t,z)\right ){\frac
{
\partial ^{2}}{\partial {t}^{2}}}\omega(x,t,z)+{\frac {\partial ^{2}}{
\partial t\partial z}}\omega(x,t,z)+\left ({\frac {\partial }{
\partial t}}\omega(x,t,z)\right )t{\frac {\partial ^{2}}{\partial t
\partial x}}\omega(x,t,z)-\]\[-\left ({\frac {\partial ^{2}}{\partial t
\partial x}}\omega(x,t,z)\right )\omega(x,t,z)
=0.
\end{equation}

   It has the particular solution defined by the
expression
\begin{equation}\label{dryuma:eq19}
\omega(x,t,z)=P(t,z)+x{e^{-z}}
\end{equation}
where the function $P(t,z)$ satisfies the equation
    \[t{e^{-z}}{\frac {\partial ^{2}}{\partial {t}^{2}}}P(t,z)+{\frac {
\partial ^{2}}{\partial t\partial z}}P(t,z)
=0,
\]
having general solution dependent from two arbitrary functions
\begin{equation}\label{dryuma:eq20}
P(t,z)={\it \_F2}(z)+\int \!{\it \_F1}(-\ln (t)-{e^{-z}}){dt} .
 \end{equation}

    In particular case

    \[{\it \_F1}(-\ln (t)-{e^{-z}})=-\ln (t)-{e^{-z}}
\]
we find the solution of the equation (\ref{dr4})
\[
f(x,y,z)={\frac {-\ln (-y{e^{z}}-{\it
\_F2}(z){e^{z}}-x){e^{z}}+z{e^{z }}-1}{{e^{z}}}}.
\]
\begin{rem}
     The equation (\ref{dryuma:eq18}) after the substitution
     \[
     \omega(x,t,z)=A(t,x-z)=A(t,\eta)
     \]
takes the form
\[
t\left ({\frac {\partial }{\partial \eta}}A(t,\eta)\right ){\frac
{
\partial ^{2}}{\partial {t}^{2}}}A(t,\eta)+t\left ({\frac {\partial }{
\partial t}}A(t,\eta)\right ){\frac {\partial ^{2}}{\partial \eta
\partial t}}A(t,\eta)-\left ({\frac {\partial ^{2}}{\partial \eta
\partial t}}A(t,\eta)\right )A(t,\eta)-{\frac {\partial ^{2}}{
\partial \eta\partial t}}A(t,\eta)=0
\]
and can be solved exactly.

   In fact, the function $B(t,\eta)=A(t,\eta)+1$ satisfies the
   equation
\[t\left ({\frac {\partial }{\partial \eta}}B(t,\eta)\right ){\frac {
\partial ^{2}}{\partial {t}^{2}}}B(t,\eta)+t\left ({\frac {\partial }{
\partial t}}B(t,\eta)\right ){\frac {\partial ^{2}}{\partial \eta
\partial t}}B(t,\eta)-\left ({\frac {\partial ^{2}}{\partial \eta
\partial t}}B(t,\eta)\right )B(t,\eta)=0
\]
which admits the first integral
\begin{equation}\label{dr8}
\left (B(t,\eta)-t{\frac {\partial }{\partial t}}B(t,\eta)\right
){ \frac {\partial }{\partial \eta}}B(t,\eta)-K(\eta)=0,
\end{equation}
 where $K(\eta)$ is arbitrary.

     Equation (\ref{dr8}) is in form
\begin{equation}\label{dr9}
\left ({\frac {\partial }{\partial y}}h(x,y)\right )h(x,y)-x\left ({
\frac {\partial }{\partial x}}h(x,y)\right ){\frac {\partial }{
\partial y}}h(x,y)-K(y)=0.
\end{equation}

    After the $(u,v)$ - transformation
    it is reduced to the relation
    \[
    \left ({\frac {\partial }{\partial t}}u(x,t)\right )u(x,t){\frac {
\partial }{\partial t}}v(x,t)-x\left ({\frac {\partial }{\partial t}}u
(x,t)\right )\left ({\frac {\partial }{\partial x}}u(x,t)\right ){
\frac {\partial }{\partial t}}v(x,t)+x\left ({\frac {\partial }{
\partial t}}u(x,t)\right )^{2}{\frac {\partial }{\partial x}}v(x,t)-\]\[-K(
v(x,t))\left ({\frac {\partial }{\partial t}}v(x,t)\right )^{2}=0
\]
which is equivalent the first order nonlinear p.d.e. with respect to the function $\omega(x,t)$
at the substitution
\[
u(x,t)=t{\frac {\partial }{\partial t}}\omega(x,t)-\omega(x,t),\quad
v(x,t)={\frac {\partial }{\partial t}}\omega(x,t)
\]
\begin{equation}\label{dr10}
-{t}^{2}{\frac {\partial }{\partial t}}\omega(x,t)+t\omega(x,t)-xt{
\frac {\partial }{\partial x}}\omega(x,t)+K\left({\frac {\partial }{
\partial t}}\omega(x,t)\right)=0.
\end{equation}

    Solutions of the equation (\ref{dr10}) depend from the function $K(\omega_t)$.

    As example in the case
    \[
    K(\omega_t)= \frac{\partial \omega(x,t)}{\partial t}
    \]

we get the linear equation
\[
-{t}^{2}{\frac {\partial }{\partial t}}\omega(x,t)+t\omega(x,t)-xt{
\frac {\partial }{\partial x}}\omega(x,t)+{\frac {\partial }{\partial
t}}\omega(x,t)=0
\]
with general solution
\[
\omega(x,t)={\it \_F1}\left({\frac {{t}^{2}-1}{{x}^{2}}}\right)x.
\]
\end{rem}
\begin{rem}
    Legendre-transformation of the equation (\ref{dr9})
\[h(x,y)=\xi\,{\frac {\partial }{\partial \xi}}\theta(\xi,\rho)+\rho\,{
\frac {\partial }{\partial \rho}}\theta(\xi,\rho)-\theta(\xi,\rho),\quad
{\frac {\partial }{\partial x}}h(x,y)=\xi,\quad
{\frac {\partial }{\partial y}}h(x,y)=\rho,\]\[
x={\frac {\partial }{\partial \xi}}\theta(\xi,\rho),\quad
y={\frac {\partial }{\partial \rho}}\theta(\xi,\rho)
\]
lead to the equation
\[
{\rho}^{2}{\frac {\partial }{\partial \rho}}\theta(\xi,\rho)-\rho\,
\theta(\xi,\rho)-K({\frac {\partial }{\partial \rho}}\theta(\xi,\rho))=0
\]
which is reduced to the Bernoulli equation after application
of the suitable Legendre transformation and so it is integrable.
 \end{rem}

   With the help of solutions of the equation (\ref{dr9}) the functions
    $A(t,\eta)$ and $ \omega(x,t,z)=A(t,x-z)=A(t,\eta)$ can be determined.

    Then after elimination of the parameter $t$ from the expressions
    for the functions $ \omega(x,t,z)$ and $y(x,t,z)$ particular solutions of the equation
    (\ref{dr4}) can be constructed.
\section{On the equations of shallow water waves}

     The equations of shallow water waves
     \begin{equation}\label{dr11}
     {\frac {\partial }{\partial y}}h(x,y)+g(x,y){\frac {\partial }{
\partial x}}h(x,y)+h(x,y){\frac {\partial }{\partial x}}g(x,y)=0,
$$$$
{\frac {\partial }{\partial y}}g(x,y)+g(x,y){\frac {\partial }{
\partial x}}g(x,y)+{\frac {\partial }{\partial x}}h(x,y)=0
\end{equation}
are intimately connected with the Benney equation.

   We apply the method of $(u,v)$-transformation for construction
   some particular solutions of the system (\ref{dr11}).

   For this purpose we use the substitution
   \[
   h(x,y)={\frac {\partial }{\partial x}}f(x,y),\quad
g(x,y)=-{\frac {{\frac {\partial }{\partial y}}f(x,y)}{{\frac {
\partial }{\partial x}}f(x,y)}}
\]
and replace the system (\ref{dr11}) by the equation with respect the function $f(x,y)$
  \begin{equation}\label{dr12}
  -\left ({\frac {\partial ^{2}}{\partial {y}^{2}}}f(x,y)\right )\left (
{\frac {\partial }{\partial x}}f(x,y)\right )^{2}+2\,\left ({\frac {
\partial }{\partial y}}f(x,y)\right )\left ({\frac {\partial ^{2}}{
\partial x\partial y}}f(x,y)\right ){\frac {\partial }{\partial x}}f(x
,y)-$$$$-\left ({\frac {\partial }{\partial y}}f(x,y)\right )^{2}{\frac {
\partial ^{2}}{\partial {x}^{2}}}f(x,y)+\left ({\frac {\partial ^{2}}{
\partial {x}^{2}}}f(x,y)\right )\left ({\frac {\partial }{\partial x}}
f(x,y)\right )^{3}=0.
\end{equation}

    The $u,v)$-transformation of the equation (\ref{dr12})
    with the conditions
    \begin{equation}\label{dr131}
    u(x,t)=t{\frac {\partial }{\partial t}}\omega(x,t)-\omega(x,t),\quad
v(x,t)={\frac {\partial }{\partial t}}\omega(x,t)
\end{equation}
 lead to the equation with respect the function $\omega(x,t)$
\begin{equation}\label{dr13}
-\left ({\frac {\partial ^{2}}{\partial {t}^{2}}}\omega(x,t)\right )
\left ({\frac {\partial ^{2}}{\partial {x}^{2}}}\omega(x,t)\right )
\left ({\frac {\partial }{\partial x}}\omega(x,t)\right )^{3}-\left ({
\frac {\partial ^{2}}{\partial {t}^{2}}}\omega(x,t)\right ){t}^{2}{
\frac {\partial ^{2}}{\partial {x}^{2}}}\omega(x,t)-$$$$-2\,t\left ({\frac
{\partial ^{2}}{\partial t\partial x}}\omega(x,t)\right ){\frac {
\partial }{\partial x}}\omega(x,t)+\left ({\frac {\partial }{\partial
x}}\omega(x,t)\right )^{2}+{t}^{2}\left ({\frac {\partial ^{2}}{
\partial t\partial x}}\omega(x,t)\right )^{2}+$$$$+\left ({\frac {\partial
^{2}}{\partial t\partial x}}\omega(x,t)\right )^{2}\left ({\frac {
\partial }{\partial x}}\omega(x,t)\right )^{3}=0.
\end{equation}

     Particular solution of the equation (\ref{dr13}) depending from
     one arbitrary function is
     \[
     \omega(x,t)=A(t)+xB(t),
\]
where the function $A(t)$ is arbitrary, and the function $B(t)$
satisfies the equation
\[
-2\,t\left ({\frac {d}{dt}}B(t)\right )B(t)+\left (B(t)\right )^{2}+{t
}^{2}\left ({\frac {d}{dt}}B(t)\right )^{2}+\left ({\frac {d}{dt}}B(t)
\right )^{2}\left (B(t)\right )^{3}=0
\]
having the solution of the form
  \begin{equation}\label{dr14}
B(t)=-\left (1/6\,\sqrt [3]{54\,t+{{\it \_C1}}^{3}+6\,\sqrt {81\,{t}^{
2}+3\,t{{\it \_C1}}^{3}}}+1/6\,{\frac {{{\it \_C1}}^{2}}{\sqrt [3]{54
\,t+{{\it \_C1}}^{3}+6\,\sqrt {81\,{t}^{2}+3\,t{{\it \_C1}}^{3}}}}}+1/
6\,{\it \_C1}\right )^{2}.
\end{equation}

    On the basis of this solution and in deciding the arbitrary function $A(t)$
we define the function $\omega(x,t)$ and then after elimination of
 the parameter $t$ from the relations (\ref{dr131}) the solutions of the
equation (\ref{dr12}) can be constructed.

    As example in particular case from (\ref{dr14}) is followed
    $$
    {\it \_C1}=0,\quad B(t)=-1/2\,\sqrt [3]{2}{t}^{2/3}
    $$
    and choosing the function $A(t)$ in the form
$$
A(t)=\sqrt [3]{t} $$  we get the relations
    \[
    6\,f(x,y){t}^{2/3}+4\,t-{t}^{4/3}\sqrt [3]{2}x=0,
\]
\[
3\,y{t}^{2/3}-1+\sqrt [3]{2}x\sqrt [3]{t}=0.
\]

   Elimination of the parameter $t$ from these relations lead to the equation for the
   function $f(x,y)$
\begin{equation}\label{dr15}
-3\,{2}^{2/3}{x}^{2}-108\,\sqrt [3]{2}xyf(x,y)+324\,{y}^{2}\left (f(x,
y)\right )^{2}-48\,y-12\,f(x,y){x}^{3}=0
\end{equation}
which is the simplest particular solution of the equation (\ref{dr12}).

   In deciding on the function $A(t)$ in more general form  can be obtained
   more complicated solutions of the equation (\ref{dr12}) and the corresponding system (\ref{dr11}).

\section{Acknowledgement}

     The research was partially supported by the Grant 06.01 CRF of HCSTD ASM.

  \end{document}